\documentclass{article}
\usepackage{spconf,amsmath,graphicx}
\usepackage{xcolor}
\usepackage{url}

\usepackage{enumitem}
\setlist{nosep, leftmargin=14pt}

\usepackage{mwe} 
\usepackage{amssymb} 
\usepackage{booktabs} 

\title{A deep learning framework for glomeruli segmentation with boundary attention
}

\name{
  \parbox{\linewidth}{\centering
    Behnaz Elhaminia\textsuperscript{a}, 
    Catherine King\textsuperscript{b,c},
    Jiaqi Lv\textsuperscript{a},
    Lorraine Harper\textsuperscript{c,e},
    Paul Moss\textsuperscript{b},\\
    Owen Cain\textsuperscript{d},
    Dimitrios Chanouzas\textsuperscript{b,c},
    Shan E Ahmed Raza\textsuperscript{a}
  }
}

\address{
  \textsuperscript{a} Tissue Image Analytics (TIA) Centre, Dept. of Computer Science, University of Warwick, UK \\
  \textsuperscript{b} Dept. of Immunology and Immunotherapy, University of Birmingham, UK \\
  \textsuperscript{c} Renal Unit, Queen Elizabeth Hospital Birmingham, UHB NHS Foundation Trust, UK \\
  \textsuperscript{d} Dept. of Cellular Pathology, Queen Elizabeth Hospital Birmingham, UK \\
  \textsuperscript{e} School of Applied Health Sciences, University of Birmingham, UK
}

\begin{document}
%
\maketitle
\begin{abstract}
Accurate detection and segmentation of glomeruli in kidney tissue are essential for diagnostic applications. Traditional deep learning methods primarily rely on semantic segmentation, which often fails to precisely delineate adjacent glomeruli. To address this challenge, we propose a novel glomerulus detection and segmentation model that emphasises boundary separation. Leveraging pathology foundation models, the proposed U-Net–based architecture incorporates a specialised attention decoder designed to highlight critical regions and improve instance-level segmentation. Experimental evaluations demonstrate that our approach surpasses state-of-the-art methods in both Dice score and Intersection over Union (IoU), indicating superior performance in glomerular delineation.
\end{abstract}
\section{Introduction}
\label{sec:intro}

The glomerulus is a key structural and functional unit of the human kidney. Accurate evaluation of glomeruli in renal biopsy samples is crucial for diagnosing a wide range of kidney diseases. This assessment also informs treatment strategies and provides valuable prognostic insights. In recent years, 
deep learning approaches have been increasingly applied to digitised whole slide images (WSIs) in diverse ways for glomerular analysis, ranging from simple combinations of classical image processing with machine learning models \cite{ginley2019computational} to the development of sophisticated deep architectures \cite{jiang2021deep}. 

Most of the recent work relies on well-known convolutional neural networks (CNNs). For instance, Moreau et al.\cite{moreau2024glomnet} proposed a detection method based on the HoverNet model, while Wang et al. \cite{wang2025glo} 
combines the Swin Transformer and the VGG module for glomerulus segmentation. Other notable contributions include the use of CNN combined with TransXNet by Lou et al.\cite{liu2024hybrid}, EfficientNetB3-Unet by He et al. \cite{he2024image}, ensembled SegNeXt by Kumar et al.\cite{kumar2024ensembled}
 and Swin Transformer combined with generative models by Yang et al. \cite{yang2025unsupervised}.

From a broad perspective, glomerulus segmentation methods can be categorised into three main approaches. Semantic segmentation aims to delineate glomerular boundaries, but these methods often cannot distinguish individual glomeruli within the same image, making it unsuitable for glomerular counting \cite{bueno2020glomerulosclerosis}. Object detection methods, on the other hand, can localise and classify glomeruli, enabling the differentiation of glomerular subtypes, but these methods do not provide precise boundary information \cite{bukowy2018region, yang2025unsupervised}. The third category, instance segmentation, detects and segments each glomerulus individually, although in many cases these models focus on identifying glomeruli rather than classifying their types \cite{jiang2021deep}. Each of these approaches faces unique challenges. One major challenge is the presence of overlapping/adjacent glomeruli, as well as glomeruli with distorted shapes and boundaries, which are difficult to delineate accurately. Another significant challenge arises from the diversity of staining protocols. In addition to hematoxylin–eosin (HE), histochemical stains such as periodic acid–Schiff (PAS) are routinely used in kidney pathology. Furthermore, pathological changes can affect different glomerular components, including endothelial cells, mesangial cells, podocytes, and the mesangial matrix, with a wide variability in both lesion type and severity \cite{jiang2021deep}. As a result, the morphological appearance of glomeruli can vary greatly between cases, making segmentation of individual glomeruli challenging.

In this study, we present a deep learning framework for glomerulus segmentation, with two primary objectives: (i) enhancing boundary delineation to enable accurate glomerular counting and detailed morphological analysis, and (ii) achieving robust detection and segmentation across diverse kidney histopathology samples, including both hematoxylin and eosin (HE) and periodic acid–Schiff (PAS) stains.  Experimental results on two datasets demonstrates the robustness and effectiveness of the proposed framework compared to state-of-the-art approaches.

\begin{figure*}[t!]
    \centering
    \includegraphics[width=0.85\textwidth]{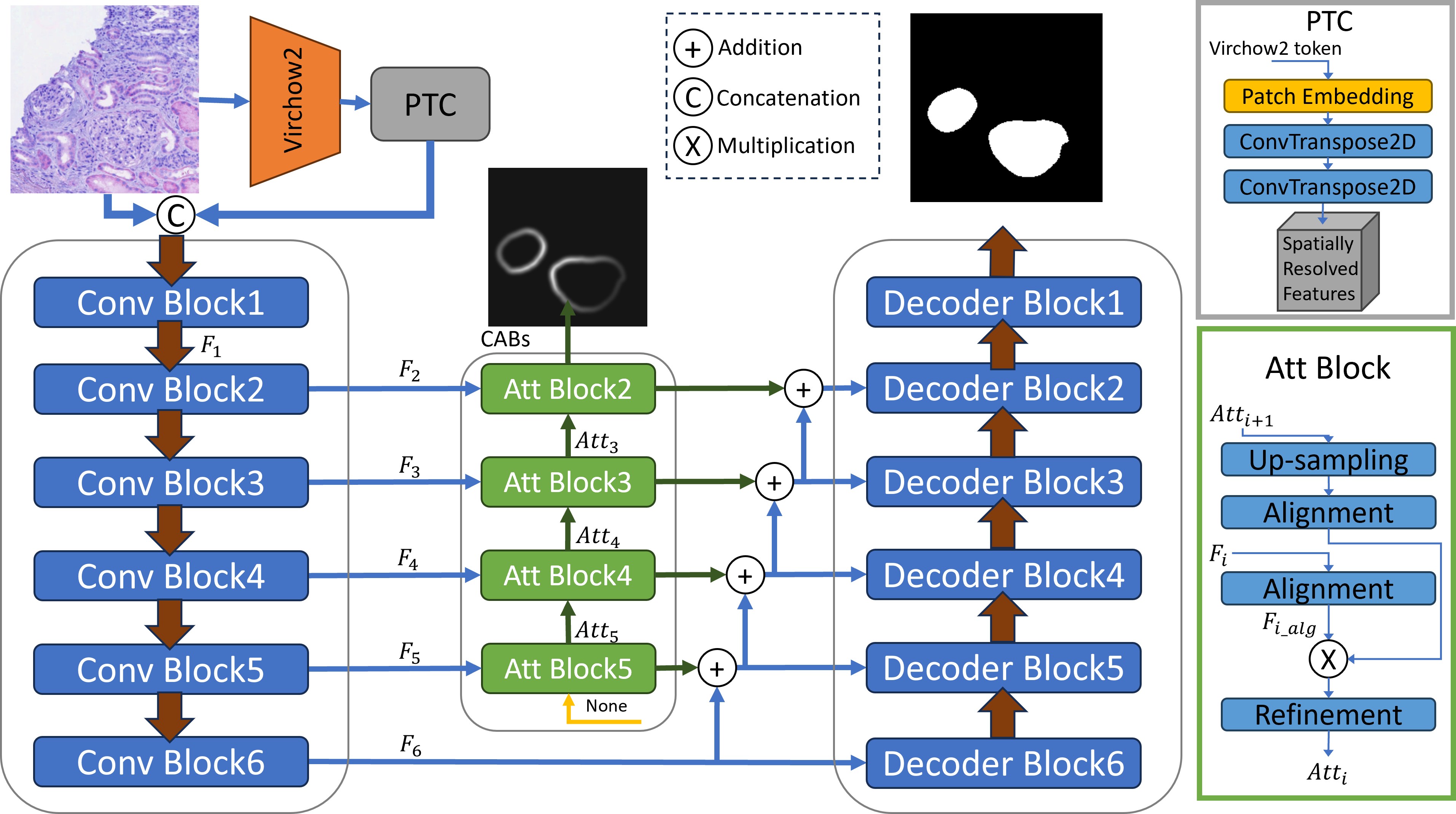}
    \caption{An illustration of the architecture and key processing stages. Deep features extracted by the Virchow2 foundation model are concatenated with the input image and fed into the encoder. The attention module exploits these features to extract critical boundary-related information, which helps the model segment boundaries with greater precision and focus.}
    \label{fig:architecture}
\end{figure*}
\section{Materials and Methods}
\label{sec:format}
\subsection{Datasets}
In this study, we used two datasets. (i) The public dataset was obtained from the Human BioMolecular Atlas Program (HuBMAP) \cite{hubmap2020} and comprises 20 PAS-stained kidney WSIs, including 11 fresh-frozen and 9 formalin-fixed paraffin-embedded (FFPE) samples. All images were scanned at a spatial resolution of 0.65 $\mu$m/pixel.
Glomerular functional tissue unit (FTU) annotations are available for all slides, providing pixel-level segmentations of glomeruli. (ii) The private dataset, was provided by University Hospitals Birmingham as part of the REACTIVAS study~\cite{hra_reactivas}. It consists of 42 renal biopsy WSIs, including 20 HE and 22 PAS-stained slides. The slides were digitised using a whole-slide scanner at $40\times$ magnification with 0.25 $\mu$m/pixel. Expert manual annotations were available for 5 HE and 3 PAS WSIs.

\subsection{Proposed model}

Our objective is to develop a deep learning model capable of segmenting various types of glomeruli across different staining protocols in renal biopsy samples, including HE and PAS stains. The model is specifically designed to improve delineation of glomerular boundaries, particularly in challenging cases where adjacent glomeruli are closely positioned. We adopt an approach similar to Lv et al. \cite{lv2025leveraging}, and employ Virchow2 as a frozen feature extractor. Using this approach, we design a novel architecture that integrates the pretrained backbone with an EfficientNetV2 \cite{tan2021efficientnetv2} based encoder-decoder into cascaded attention modules. Furthermore, we introduce a new adaptive weighted loss function to improve boundary delineation and segmentation accuracy in challenging regions.

The proposed architecture is illustrated in Fig. ~\ref{fig:architecture}. For an input image $I \in \mathbb{R}^{W\times H\times3}$, we concatenate the image with spatially resolved features extracted from Virchow2 combined with the PTC module~\cite{lv2025leveraging}. The concatenated features are processed by the encoder to generate multi-scale representations $F_i$, where $i \in \{1,2,3,4,5,6\}$.

\textbf{Cascaded Attention Blocks:} To enhance boundary localisation, we introduce Cascaded Attention Blocks (CABs) designed to focus on glomerular boundaries. 
Previous studies have shown that incorporating boundary information, such as the gradient map of a binary ground truth, can improve segmentation accuracy \cite{zhu2022can,chen2018reverse}. We build on the concept introduced by Zhu et al.~\cite{zhu2022can}. However, instead of emphasising foreground–background transitions, we adaptively focus on challenging boundary regions using learnable parameters. The module receives four feature maps from the encoder, in order, from deep (low-resolution) to shallow (high-resolution) layers. Each block consists of two components: (i) Alignment and Refinement, which adjusts channel dimensions and refines fused features, and (ii) Up-sampling, which increases spatial resolution via bilinear interpolation.


The encoder features \( F_i \) from layer \( i \) are fed into CABs, which function as a decoder to generate critical boundary weights. To reduce complexity, we utilise encoder features from layers 2 to 5, excluding the shallowest and deepest layers that are less informative for boundary localisation. The attention feature for layer \( i \) is computed as:
\[
Att_i = \mathrm{Conv}\big(\mathrm{Conv}(F_i) \odot \mathrm{Conv}(\mathrm{Up}(A_{i+1}))\big),
\]
where $A_{i+1}$ is the attention features for deeper layer $i+1$, \( \odot \) denotes element-wise multiplication, \( \mathrm{Up}(\cdot) \) represents upsampling, and \( \mathrm{Conv}(\cdot) \) indicates a  convolution block. The final attention block processes only the feature \( F_5 \) using a single convolutional block. This setup integrates boundary cues into the feature space, improving model sensitivity to fine structures. CABs are trained in a supervised manner using boundary weight maps derived from glomerular masks (see \textbf{Boundary weighted Loss Fucntion}) as ground truth. The attention module functions as a compact decoder that highlights critical boundary regions. The final output of the CABs is obtained using Softmax, producing a probability map emphasising most challenging boundary regions.
\begin{figure}[t!]
    \centering    \includegraphics[width=0.49\textwidth]{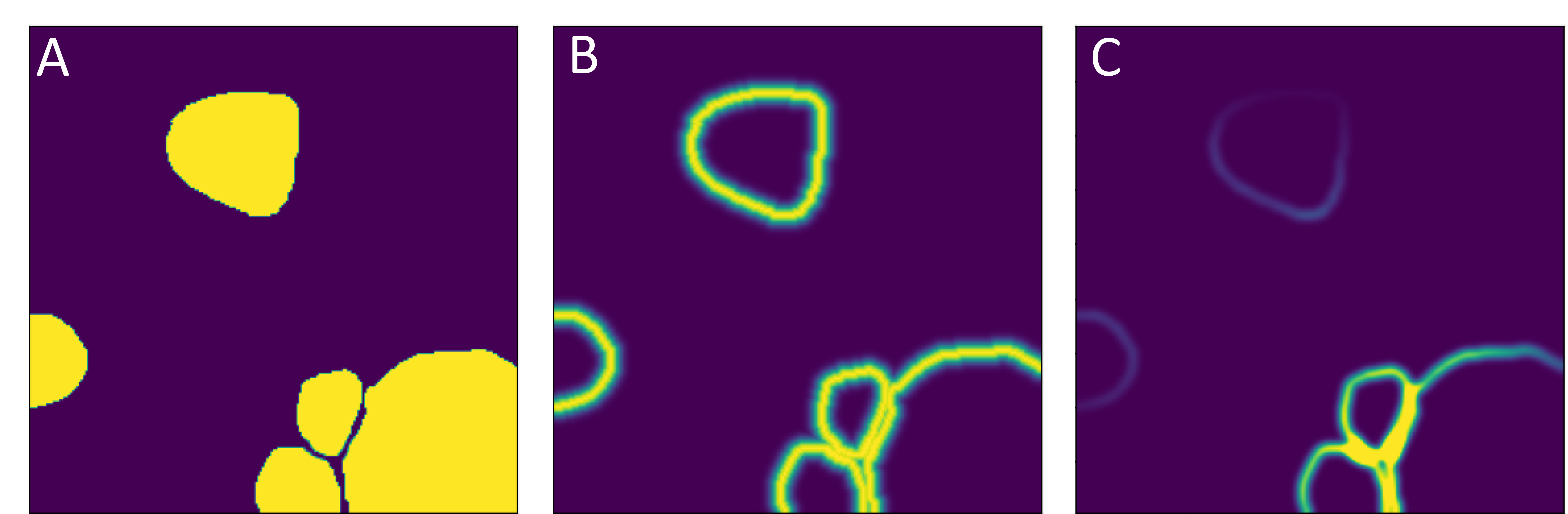}
    \caption{Comparison of static and adaptive weight maps. (A) Glomeruli mask, (B) Static distance-transform weight map, and (C) Adaptive boundary weighting map. The adaptive weights assign higher importance to challenging boundaries between closely adjacent glomeruli.}
    \label{fig:adaptive_weights}
\end{figure}

\textbf{Boundary weighted Loss Function:} Previous studies have used weighted loss based on distance transform to separate the boundaries. However, these approaches treat all borders equally. We propose a dynamic boundary weighting that adaptively emphasises challenging regions, increasing weights where the glomeruli are closely spaced to guide the model for challenging boundaries. Mathematically, the weight map $W(x)$ is defined as: $W(x) = w_0 \cdot \exp\left(-{D(x)^2}/{2\sigma^2}\right)$,
where $D(x)$ denotes the minimum distance from a boundary pixel $x$ to the nearest boundary of another glomerulus, $w_0$ is a scaling factor, and $\sigma$ controls the rate of exponential decay. This ensures that closely packed glomeruli receive greater emphasis during training, while all object borders retain meaningful weights. Fig. \ref{fig:adaptive_weights} illustrates the results for adaptive weights compared to static weights. These weight maps are also used as a ground truth for training CABs. The final training loss combines a weighted binary cross-entropy (WBCE) loss, $\mathcal{L}_{\text{WBCE}}$~\cite{ronneberger2015u}, and a Dice loss $\mathcal{L}_{\text{Dice}}$~\cite{lv2025leveraging} for the encoder-decoder, along with an attention loss as: 
 \begin{equation}
\mathcal{L} = \gamma\mathcal{L}_{\text{Att}} + \mathcal{L}_{\text{WBCE}} + \mathcal{L}_{\text{Dice}},
\end{equation}
The attention loss $\mathcal{L}_{\text{Att}}$ is defined as
\[
\mathcal{L}_{\text{Att}} = - \frac{1}{N} \sum_{i=1}^{N} \big[ W_i \log(A_i) + (1 - W_i) \log(1 - A_i) \big],
\]
where \(A_i \in [0,1]\) is the predicted attention map value for pixel \(i\), \(W_i \in [0,1]\) is the corresponding boundary weight map used as ground truth, and \(N\) is the total number of pixels. The weighting factor $\gamma$ was tuned during hyperparameter optimisation to a value of 0.6. 

\section{Experimental Results}
\label{sec:experiments}

We evaluated our method in comparison with the top three models on the Kaggle HuBMAP challenge leaderboard~\cite{jain2023segmentation}. The first model, TOM architecture~\cite{hubmap_github}, employs a U-Net–style design with a SeResNeXt101 encoder and an attention module. The second model, Gleb~\cite{jain2023segmentation}, uses an ensemble of four four-fold models, including three U-Net–style networks with different encoders and an attention-based decoder. The third model, What’s Going On (WGO)~\cite{he2025_hubmap3rdplace}, is an ensemble of two U-Net–style five-fold models. A Feature Pyramid Network (FPN) is incorporated to provide skip connections between the upscaling blocks of the decoder.

To further investigate the effectiveness of the proposed attention module and boundary weighting strategy, we also compared our method with the work by Lv et al. \cite{lv2025leveraging}, which employs the same backbone but excludes the cascaded attention module. Their work was originally proposed for tissue segmentation of five tissue classes; we adapted the architecture for our two-class problem and refer to it as Eff-UNet.

For comparison, we evaluate the models in terms of the Dice and Intersection over Union (IoU) metrics \cite{maier2024metrics}. All models were trained using five-fold cross-validation on image patches extracted from both the HuBMAP and REACTIVAS datasets. Testing was performed on unseen data from the respective test sets. The HuBMAP training and validation set consisted of 15 WSIs, with evaluation conducted on five WSIs. In contrast, the REACTIVAS dataset included 2,977 HE and PAS patches in the training and validation set, with testing performed on 676 HE and 471 PAS unseen patches. All test patches contained foreground tissue regions to ensure a meaningful assessment. Furthermore, for model generalisation, we applied strong data augmentations during training. These included random RGB shifts, image compression, Gaussian blur and sharpening, random hue–saturation–value (HSV) shifts, random translations and scaling, brightness and contrast adjustments, and random rotations with horizontal and vertical flips. The official training and inference codes are publicly available for TOM and WGO. However, the training code for Gleb is not available; therefore, we only report the Dice score results on the HuBMAP dataset for Gleb. The performance comparison is summarised in Table~\ref{tab:comparison1} and Table~\ref{tab:comparison2}. 
\begin{table}[t!]
    \centering
    \caption{Performance Comparison across two datasets in terms of Dice scores ($\mu \pm \sigma$). For REACTIVAS, we add the results for HE and PAS separately for further insights. Training code is not available for Gleb. Therefore, we could not train Gleb on the REACTIVAS dataset.\vspace{6pt}}
    \label{tab:comparison1}
    \setlength{\tabcolsep}{3pt} 
    \begin{tabular}{l c c c c}
        \toprule
        & \textbf{HubMap} & \multicolumn{3}{c}{\textbf{REACTIVAS}} \\
        \cmidrule(lr){3-5} 
        \textbf{Method} & \textbf{PAS} & \textbf{HE} & \textbf{PAS} & \textbf{All} \\
        \midrule
        Proposed &  \textbf{0.9630} & $ \textbf{0.89}\pm \textbf{0.3}$ & $ \textbf{0.86}\pm \textbf{0.1}$ & \textbf{0.88} $\pm$ \textbf{0.01} \\
        Eff\text{-}UNet & 0.9111 & $ 0.84\pm 0.2$ & $ 0.83\pm 0.4$ & $ 0.84\pm 0.3$ \\
        Tom & 0.9515 & $ 0.80\pm 0.1$ & $ 0.82\pm 0.4$ & $ 0.81\pm 0.2$ \\
        Gleb & 0.9507 & - & - & - \\
        WGO &  0.9503 & $ 0.79\pm 0.2$ & $ 0.81\pm 0.3$ & $ 0.80\pm 0.2$ \\
        \bottomrule
    \end{tabular}
\end{table}
\begin{table}[t!]
    \centering
    \caption{Performance Comparison across two datasets in terms of IoU ($\mu \pm \sigma$). For REACTIVAS, we add the results for HE and PAS separately for further insights. IoU results were not available for the Gleb model.\vspace{6pt}}
    \label{tab:comparison2}
    \setlength{\tabcolsep}{3pt} 
    \begin{tabular}{l c c c c}
        \toprule
        & \textbf{HubMap} & \multicolumn{3}{c}{\textbf{REACTIVAS}} \\
        \cmidrule(lr){3-5} 
        \textbf{Method} & \textbf{PAS} & \textbf{HE} & \textbf{PAS} & \textbf{All} \\
        \midrule
        Proposed & 0.941 & \textbf{0.88$\pm$0.2}  & \textbf{0.86$\pm$ 0.3} & \textbf{0.87}$\pm$\textbf{0.4} \\
        Eff\text{-}UNet & 0.904 & $0.82\pm$0.1 & 0.81$\pm$ 0.1& 0.82$\pm$0.3 \\
        Tom & \textbf{0.943} & 0.78$\pm$0.2 & 0.79$\pm$ 0.4& 0.78$\pm$0.3 \\
        WGO &  0.931& 0.77$\pm$0.2 &0.79 $\pm$0.1 & 0.78$\pm$0.6 \\
        \bottomrule
    \end{tabular}
\end{table}
\begin{figure*}[t!]
    \centering
    \includegraphics[width=0.99\textwidth]{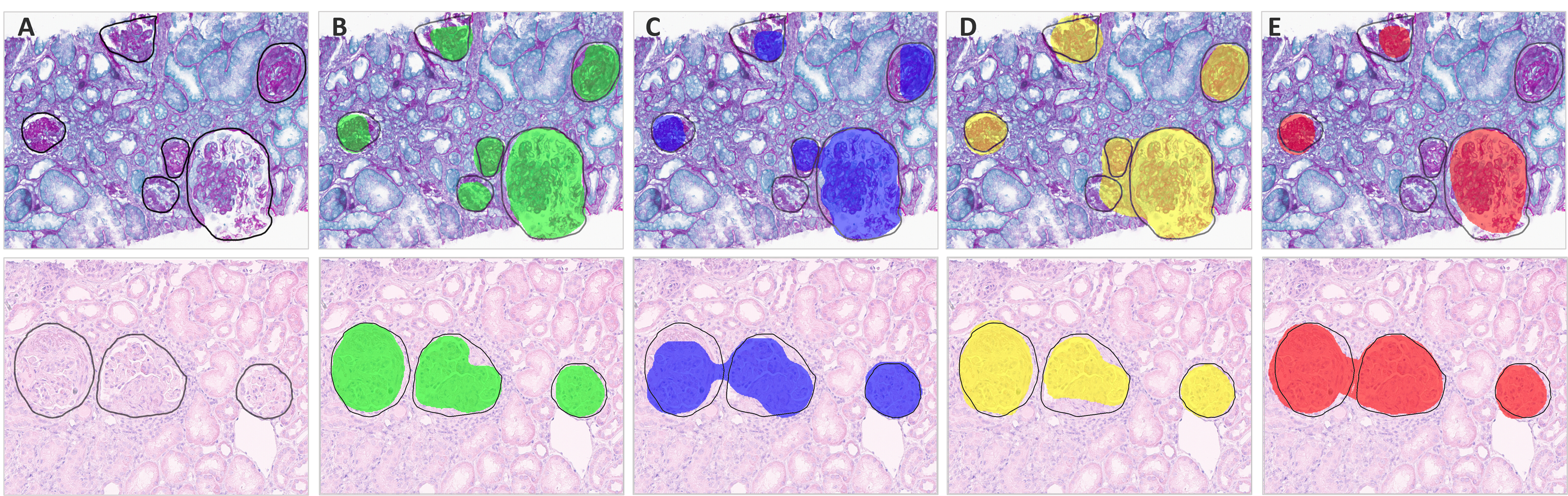}
    \caption{Comparison of glomerulus segmentation results, with PAS images in the top row and HE images in the bottom row. (A) Ground truth, (B) Proposed model, (C) Eff-UNet, (D) TOM, and (E) WGO. The patches are taken from the test set of the REACTIVAS dataset and illustrate challenging cases with closely spaced glomeruli.}
    \label{fig:imageComaprison}
\end{figure*} 

For qualitative evaluation, we provide representative patches from the REACTIVAS dataset in Fig.~\ref{fig:imageComaprison}, which shows superior performance of the proposed model for segmenting neighbouring glomeruli.

\section{Conclusion}
\label{sec:conclusion}

The results in Table~\ref{tab:comparison1} and Table~\ref{tab:comparison2} show that the Dice and IoU decreased for all models when tested on the REACTIVAS dataset. This is likely due to the increased complexity of the REACTIVAS data closely adjacent glomeruli sharing borders, making segmentation more difficult. Despite these challenges, the proposed model outperformed state-of-the-art due to its attention to boundaries. Apart from the CABs module, the proposed boundary-weighted loss mechanism allows the loss function to dynamically adapt to the local spatial configuration of glomeruli, promoting precise boundary delineation and reducing the likelihood of merging adjacent structures. Direct comparison with Eff-UNet highlights the effectiveness of the CABs module for segmentation for both Dice and IoU scores. 
{While the results are encouraging, this study is limited by the small test sets, which may affect generalisability. We plan to validate the model on larger datasets and conduct ablation studies to assess the contribution of key components, including the CABs, adaptive weighting, and Virchow2 features, to better understand their impact on performance.}

In conclusion, glomeruli segmentation benefits from identifying critical regions rather than treating all borders equally. Focussing attention and assigning higher weights to these regions can improve detection and boundary delineation.


\section{Acknowledgments}
\label{sec:acknowledgments}
BE \& SR report financial support from MRC (MR/X011585/1). CK is funded by an MRC CRTF fellowship (MR/X006964/1) which supported the collection of the REACTIVAS dataset. Recruitment of patients to the REACTIVAS dataset was partly supported by an Investigator-Initiated Program of Merck Sharp a Dohme Corp (MSD). JL is supported by the EPSRC UK. 


\begingroup
\bibliographystyle{IEEEbib}
\bibliography{strings,refs}
\endgroup


\end{document}